# Parallel Computing for Copula Parameter Estimation with Big Data: A Simulation Study

Zheng Wei[1], Daeyoung Kim[1], Erin M. Conlon[1*]

[1] Department of Mathematics and Statistics, University of Massachusetts, Amherst, Massachusetts, United States of America

* Corresponding author

E-mail: econlon@mathstat.umass.edu

# Abstract

Copula-based modeling has seen rapid advances in recent years. However, in big data applications, the lengthy computation time for estimating copula parameters is a major difficulty. Here, we develop a novel method to speed computation time in estimating copula parameters, using communication-free parallel computing. Our procedure partitions full data sets into disjoint independent subsets, performs copula parameter estimation on the subsets, and combines the results to produce an approximation to the full data copula parameter. We show in simulation studies that the computation time is greatly reduced through our method, using three well-known one-parameter bivariate copulas within the elliptical and Archimedean families: Gaussian, Frank and Gumbel. In addition, our simulation studies find small values for estimated bias, estimated mean squared error, and estimated relative $L_1$ and $L_2$ errors for our method, when compared to the full data parameter estimates.

# 1. Introduction

The field of copula-based modeling has grown rapidly in recent years for estimating multivariate dependence. Copulas are used in a broad range of subject areas including finance (Nikoloulopoulos et al., 2012; Fang & Madsen, 2013; Rodriguez, 2007; Junker & May, 2005; Cherubini et al., 2004), econometrics (Patton, 2006; Sancetta & Satchell 2004; Fermanian & Scaillet, 2003), biology and medicine (Winkelmann, 2012; de Leon & Wu, 2011; Kim et al., 2008), environmental science and engineering (Zhang & Singh, 2007; Yan 2006, Genest & Favre, 2007), and actuarial science (Otani & Imai, 2013; Frees & Wang, 2005; Frees & Valdez, 1998). Their wide use is due in part to their flexibility in modeling; the primary feature of copulas is that the dependence structure of random variables can be specified separately from their marginal distributions, when constructing joint distributions. However, for large data sets, copula parameter estimation can be challenging and often requires excessive computing time (Emrouznejad, 2016; Cevher et al., 2014; Slavakis et al., 2014). In addition, it may not be possible to analyze big data sets in full, either due to limits on computer memory or storage capacity, with data stored on multiple machines.

As a result of the difficulties in estimating copula parameters for big data, we introduce a new method for estimating copula parameters that involves parallel computing. Our procedure partitions full data sets into disjoint independent subsets, analyzes all subsets separately and combines results to estimate full data copula parameters. Note that the estimates are based on all data values in a full data set, since the subsets are a partition of the full data set. Since each subset is analyzed independently of other subsets, with no communication of results between subsets, this is referred to as communication-free parallel computing (or embarrassingly parallel;

see Foster, 1995). This is in contrast to parallel computing tasks that require communication between processes (or machines). For this, each process must contact other processes to exchange intermediate data, and each process depends on others to conclude their computations; examples include image analysis applications and heat conduction applications (Mighell, 2010; Zahid et al., 2011).

Here, we illustrate our new method using bivariate distributions with three commonly-used copula models, Gaussian, Frank and Gumbel; each of these copulas has one parameter. Simulation studies demonstrate that our new parallel computing technique greatly reduces computation time, and produces full data copula parameter estimates with small estimated bias, small estimated mean squared error (MSE) and small estimated relative $L_1$ and $L_2$ errors.

Our paper is organized as follows. In Section 2, we provide a brief introduction to copula theory, and introduce our new parallel computing method. In Section 3, we describe the design of the simulation studies and show results for the bivariate Gaussian, Frank and Gumbel copulas. Our overall findings are summarized in Section 4.

## 2 Methods

### 2.1 Background on copula theory

Copulas provide a method for creating joint distributions by specifying marginal distributions and their dependence structure separately. Here, we briefly review copula theory; more general copula theory can be found in Joe (1997) and Nelsen (2006).

<u>Copula definition:</u> A *copula* is the joint distribution of $k$ random variables $U_1, U_2, ..., U_k$, each of which has a marginal uniform distribution $U(0,1)$. Alternatively, a *copula* is the joint cumulative distribution function of the above distribution:

$$C(u_1, u_2, ..., u_k) = P(U_1 \leq u_1, U_2 \leq u_2, ..., U_k \leq u_k). \tag{1}$$

Copulas are useful due to Sklar's Theorem (Sklar, 1959), which states that a $k$-dimensional joint distribution can be separated into $k$ univariate marginal distributions and a $k$-dimensional copula, i.e.:

<u>Sklar's theorem:</u> For any $k$ random variables $X_1, X_2, ..., X_k$, that have the joint cumulative distribution function

$$F(x_1, x_2, ..., x_k) = P(X_1 \leq x_1, X_2 \leq x_2, ..., X_k \leq x_k) \tag{2}$$

with marginal cumulative distribution functions

$$F_j(x) = P(X_j \leq x), \quad j = 1, 2, ..., k, \tag{3}$$

there exists a copula $C$ such that

$$F(x_1, x_2, ..., x_k) = C[F_1(x_1), F_2(x_2), ..., F_k(x_k)]. \tag{4}$$

Note that if each $F_j(x)$ is continuous, then $C$ is unique.

The probability density function $c$ of the copula distribution is obtained by the following, when $F(\bullet)$ and $C(\bullet)$ are differentiable (Joe, 1997; Nelsen, 2006):

$$c(u_1, u_2, ..., u_k) = \frac{\partial^k}{\partial u_1 \partial u_2 ... \partial u_k} C(u_1, u_2, ..., u_k). \tag{5}$$

## 2.2 Families of copulas

Here, we describe the two most frequently used classes of copulas, elliptical and Archimedean, as well as the specific copulas within these classes, Gaussian, Frank and Gumbel. For a guide to copula modeling, see Trivedi & Zimmer (2005).

**2.2.1 Elliptical copulas.** Elliptical copulas are the copulas of elliptical distributions, and are widely used in econometrics and finance in particular (Frahm et al., 2003; Fang et al., 1987; Cambanis et al., 1981). Elliptical copulas are defined by Sklar's theorem above, with an elliptical cumulative distribution function $F$, as follows (Yan, 2007; Fang et al., 1987). Suppose $F_i$ is the cumulative distribution function of the $i$th marginal and its inverse is $F_i^{-1}$, for $i = 1,...,k$. Then the elliptical copula derived by $F$ is

$$C(u_1, u_2, ..., u_k) = F\left[F_1^{-1}(u_1), ..., F_k^{-1}(u_k)\right]. \tag{6}$$

The Gaussian copula is the most commonly-used special case of the elliptical copula (Pitt et al., 2006; Malevergne & Sornette, 2003; Song, 2000). If $F(\bullet)$ is the $k$-dimensional multivariate normal distribution with mean vector $\boldsymbol{\mu}$ and covariance matrix $\boldsymbol{\Sigma}$, denoted by $N_k(\boldsymbol{\mu}, \boldsymbol{\Sigma})$, then $C(\bullet)$ is a Gaussian copula. The Gaussian copula does not change when the location or scale of $N_k(\boldsymbol{\mu}, \boldsymbol{\Sigma})$ is changed; therefore, typically $\boldsymbol{\mu} = \boldsymbol{0}$ and $\boldsymbol{\Sigma} = \boldsymbol{R}$, where $\boldsymbol{R}$ is a correlation matrix. Specifically, the *k*-dimensional Gaussian copula is defined by the following:

$$C_k(u_1, u_2, ..., u_k; \boldsymbol{R}) = \Phi_k(\Phi^{-1}(u_1), \Phi^{-1}(u_2), ..., \Phi^{-1}(u_k); \boldsymbol{R}), \tag{7}$$

where $\Phi(\bullet)$ is the standard normal cumulative distribution function, and $\Phi^{-1}(\bullet)$ is the corresponding quantile function. For the bivariate distribution, the Gaussian copula *C* can be written as (Huard et al., 2006; Embrechts et al., 2003)

$$C_{Gaussian}(u_1, u_2; \theta) = \int_{-\infty}^{\Phi^{-1}(u_1)} \int_{-\infty}^{\Phi^{-1}(u_2)} \frac{1}{2\pi(1-\theta^2)^{1/2}} \exp\left\{-\frac{s^2 - 2\theta st + t^2}{2(1-\theta^2)}\right\} ds\, dt, \quad -1 \leq \theta \leq 1. \tag{8}$$

Here, $\theta$ is the dependence parameter, and is the parameter to be estimated for this copula.

**2.2.2 Archimedean copulas.** The Archimedean copulas are extensively used in applications such as finance and insurance, as well as biological subjects such as survival analysis (Prenen et al., 2014; Embrechts et al., 2003; Bouye et al., 2000). The wide appeal of the most commonly-used Archimedean copulas is due in part to their simple closed-form expressions. This is in contrast to the elliptical copulas, for which no closed form expressions exist. Archimedean copulas also allow for diverse dependence structures among random variables (Embrechts et al., 2003). An Archimedean copula is created by a generator $\psi$ through

$$C(u_1, u_2, ..., u_k) = \psi\left(\psi^{-1}(u_1) + \psi^{-1}(u_2) + \cdots + \psi^{-1}(u_k)\right), \tag{9}$$

where $\psi^{-1}$ is the inverse of the generator (Genest & MacKay, 1986; Nelsen, 2006; Yan, 2007).

A prevalent Archimedean copula is the Frank copula, which is a symmetric copula with strong dependence in the middle of the distribution, and weak dependence in the tails. In addition, the tails of the distribution are lighter than the Gaussian copula (Venter, 2002). The bivariate Frank copula is given by (Genest et al., 2009; Huard et al., 2006; Weiss, 2011)

$$C_{Frank}(u_1, u_2; \theta) = -\frac{1}{\theta} \ln\left(1 + \frac{(e^{-\theta u_1} - 1)(e^{-\theta u_2} - 1)}{e^{-\theta} - 1}\right), \quad \theta \in \mathbb{R} \setminus \{0\}. \tag{10}$$

Here, $\theta$ is the dependence parameter, and is the parameter to be estimated for this copula.

Another frequently-used Archimedean copula is the Gumbel-Hougaard copula; this is appropriate for modeling upper-tail dependence. In the bivariate case, it is referred to as the Gumbel copula, and has strong right-tail dependence with relatively weak left-tail dependence. It is given by (Joe, 1997; Genest et al., 2009)

$$C_{Gumbel}(u_1, u_2; \theta) = \exp\left(-\left[(-\ln u_1)^\theta + (-\ln u_2)^\theta\right]^{1/\theta}\right), \quad \theta \geq 1, \tag{11}$$

where $\theta$ is again the dependence parameter, and is the parameter to be estimated for this copula.

## 2.3 Methods for estimating parameters in copula models

There are many methods for estimating copula parameters for a given data sample, including the fully parametric maximum likelihood method (Yan, 2007), the semiparametric maximum pseudo-likelihood (MPL) method (Genest et al., 2009), the inference functions for margins method (Joe, 2014), the inversion of Kendall's tau estimator (Yan, 2007), and the inversion of Spearman's rho estimator (Yan, 2007). Here, we focus on the MPL method, since it has outperformed other methods in several studies (Kim et al., 2007; Tsukahara, 2005); it is also the default method in the R package **copula** (Yan, 2007; R Core Team). The MPL method is similar to the maximum likelihood method, except that the marginal distributions are first replaced with their normalized ranks; the MPL method for the data matrix $X$ is described in detail below (Genest et al., 2009; Oakes, 1994; Genest et al., 1995; Shih & Louis, 1995). We first define notation for the data matrix $X$ for a sample of size $N$ by the following. Let $x_1, x_2, ..., x_N$ be a sample of size $N$ from a continuous random vector $(X_1, ..., X_k)^T$, where each $x_i$, $i = 1, ..., N$, is a $k \times 1$ vector, and $k$ is the number of continuous random variables associated with the copula. Then the data matrix $X$ of dimension $N \times k$ is given by $X = (x_1, x_2, ..., x_N)^T$. The MPL method for $X$ is described next.

MPL method: The first step in the MPL method is to define the ranks for the data matrix $X$ as follows, with $X$ as defined above. For the $j$th marginal random variable, $X_j$, $r_{ij}$ is the rank of $x_{ij}$ for the $N$ data points from $X_j$, $x_{1j}, ..., x_{Nj}$. The normalized ranks $u_{ij}$ for $X_j$ are then determined by

$$u_{ij} = \frac{r_{ij}}{N+1}; \quad i = 1, ..., N; \, j = 1, ..., k. \tag{12}$$

For the copula with parameter $\theta$, the pseudo log-likelihood function is then represented as

$$l_{PSEUDO}(\theta) = \log \prod_{i=1}^{N} c(u_{i1}, u_{i2}, ..., u_{ik}; \theta) = \sum_{i=1}^{N} \log c(u_{i1}, u_{i2}, ..., u_{ik}; \theta). \tag{13}$$

Here, $c(u_1, u_2,...,u_k; \theta)$ is the density of the copula associated with the random vector

$(X_1,..., X_k)^T$. The MPL estimators are produced by maximizing Equation (13) with respect to $\theta$ using numerical methods (Genest et al., 2009). Calculation of the asymptotic variance of the MPL estimators requires computation of partial derivatives of the log-copula density. The asymptotic variance can then be created numerically according to the methods described in Genest et al. (1995); see also Kojadinovic & Yan (2010).

In the simulation studies below, we estimate the copula parameters using the MPL method for both the subsets and the full data analyses, and the copula model is assumed to be known. We also calculate the asymptotic variance of the MPL estimators as described above; all estimates are carried out using the R package **copula** (Yan, 2007; R Core Team, 2016).

## 2.4 New parallel computing method for estimating copula parameters

Here, we introduce a new method for copula parameter estimation that combines subset results from communication-free parallel computing; this procedure is for single-parameter bivariate copulas. As an overview, the full data set is partitioned into disjoint independent subsets, the subsets are analyzed, and the results are combined to estimate the full data parameter. More specifically, the full data matrix $X = (x_1, x_2,..., x_N)^T$ of dimension $N \times k$ (with $X$ as defined in Section 2.3) is partitioned into $M$ disjoint independent subsets $X_m$, $m = 1,…,M$. The partitioning is by the rows, so that if $X$ has dimension $N \times k$, then $X$ is partitioned as follows:

$$X = \begin{pmatrix} X_1 \\ X_2 \\ \vdots \\ X_M \end{pmatrix}; \tag{14}$$

here, each $X_m$, $m = 1,...,M$, has $k$ columns (where, again, $k$ is the number of continuous random variables associated with the copula). For each independent subset analysis, the unknown copula parameter is estimated through the MPL procedure as specified in Section 2.3. We then combine the subset parameter estimates through a weighted average, where the weights are the inverses of the asymptotic variances of the subset copula parameter estimates (using the asymptotic variance calculation method described in Section 2.3). In notational summary, the combined estimated copula parameter $\hat{\hat{\theta}}_{Combined}$ is calculated for $M$ subsets by the following:

$$\hat{\hat{\theta}}_{Combined} = \frac{\sum_{m=1}^{M} w_m \hat{\theta}_m}{\sum_{m=1}^{M} w_m}; \text{ where } w_m = \frac{1}{\sigma_m^2}; \ m = 1,...,M. \tag{15}$$

Here, $\hat{\theta}_m$ is the estimated copula parameter in subset $m$, and $\sigma_m^2$ is the asymptotic variance of $\hat{\theta}_m$. We use the "double hat" notation for $\hat{\hat{\theta}}_{Combined}$, since it is an estimate of the estimate $\hat{\theta}_{Full}$, which is based on the full data set. Note that this method requires that the asymptotic variance of the copula parameter can be estimated.

## 2.5 Simulation study design

For each of the copula models, Gaussian, Frank and Gumbel, data was simulated from the specific copula $C$ as defined in Section 2.2, with the following parameters. For the Gaussian copula, $\theta_{Gaussian} = 0.3$ (Spearman's rho = 0.29; Kendall's tau = 0.19); for the Frank copula, $\theta_{Frank} = 5$ (Spearman's rho = 0.64; Kendall's tau = 0.46); and for the Gumbel copula, $\theta_{Gumbel} = 5$ (Spearman's rho = 0.94; Kendall's tau = 0.80). The full data sets were generated with sample sizes of $N = 50{,}000$; $100{,}000$; and $200{,}000$, so that the full data analyses were still achievable. The number of subsets was set to $M = 10$, $20$ and $100$ for each $N$, with each subset having sample size $N/M$. We repeated the simulation procedure for a total of $S = 50$ times, for each combination of $N$ and $M$ for each copula.

For the repeated simulation studies, the combined estimated copula parameter $\hat{\hat{\theta}}_{Combined,s}$ is calculated for $M$ subsets and each simulation study $s = 1,\ldots,S$ by the following:

$$\hat{\hat{\theta}}_{Combined,s} = \frac{\sum_{m=1}^{M} w_{m,s} \hat{\theta}_{m,s}}{\sum_{m=1}^{M} w_{m,s}}; \text{ where } w_{m,s} = \frac{1}{\sigma_{m,s}^2}; \; m = 1,\ldots,M. \tag{16}$$

Here, $\hat{\theta}_{m,s}$ is the estimated copula parameter in subset $m$ of simulation study $s$, and $\sigma_{m,s}^2$ is the asymptotic variance of $\hat{\theta}_{m,s}$. The final combined estimated copula parameter over all simulation studies is

$$\hat{\bar{\theta}}_{Combined(Sim)} = \frac{\sum_{s=1}^{S} \hat{\bar{\theta}}_{Combined,s}}{S}. \tag{17}$$

For estimates based on full data sets, the final estimated copula parameter over all simulation studies is calculated as

$$\hat{\theta}_{Full(Sim)} = \frac{\sum_{s=1}^{S} \hat{\theta}_{Full,s}}{S}, \tag{18}$$

where $\hat{\theta}_{Full,s}$ is the estimated full data copula parameter of simulation study $s$, $s=1,\ldots,S$.

## 2.6 Metrics for comparing full data analysis versus combined subset analysis

We evaluate our new method using estimated bias, estimated MSE, and estimated relative $L_1$ and $L_2$ errors for the simulation studies. For these values, we compare the results of our method versus estimates that would be obtained by a full data analysis. The following four definitions are used (Weiss, 2011; Neiswanger et al., 2014; Oliva et al., 2013):

1) $\widehat{\text{Bias}}\left(\hat{\bar{\theta}}_{Combined(Sim)}\right) = \hat{\bar{\theta}}_{Combined(Sim)} - \hat{\theta}_{Full(Sim)}$.

2) $\widehat{\text{MSE}}\left(\hat{\bar{\theta}}_{Combined(Sim)}\right) = \dfrac{\sum_{s=1}^{S}(\hat{\bar{\theta}}_{Combined,s} - \hat{\theta}_{Full,s})^2}{S}$, where $s$ is the simulation study, $s = 1,\ldots,S$.

3) The estimated relative $L_1$ distance, $d_{1,\text{Relative}}(\hat{c},\hat{\hat{c}})$, between the estimated joint density $\hat{c}$ of the copula distribution with the parameter estimated from the full data set and the estimated joint density $\hat{\hat{c}}$ with the parameter estimated from our subset combining method is defined as follows:

$$d_{1,\text{Relative}}(\hat{c},\hat{\hat{c}}) = \frac{\|\hat{c}-\hat{\hat{c}}\|_{L_1}}{\|\hat{c}\|_{L_1}} = \frac{\int_0^1\int_0^1 \left|\hat{c}(u_1,u_2;\hat{\theta}_{Full(Sim)}) - \hat{\hat{c}}(u_1,u_2;\hat{\hat{\theta}}_{Combined(Sim)})\right| du_1 du_2}{\int_0^1\int_0^1 \left|\hat{c}(u_1,u_2;\hat{\theta}_{Full(Sim)})\right| du_1 du_2}. \quad (19)$$

4) The estimated relative $L_2$ distance, $d_{2,\text{Relative}}(\hat{c},\hat{\hat{c}})$, is calculated similarly to $L_1$, as follows:

$$d_{2,\text{Relative}}(\hat{c},\hat{\hat{c}}) = \frac{\|\hat{c}-\hat{\hat{c}}\|_{L_2}}{\|\hat{c}\|_{L_2}} = \frac{\left(\int_0^1\int_0^1 \left(\hat{c}(u_1,u_2;\hat{\theta}_{Full(Sim)}) - \hat{\hat{c}}(u_1,u_2;\hat{\hat{\theta}}_{Combined(Sim)})\right)^2 du_1 du_2\right)^{1/2}}{\left(\int_0^1\int_0^1 \left(\hat{c}(u_1,u_2;\hat{\theta}_{Full(Sim)})\right)^2 du_1 du_2\right)^{1/2}}. \quad (20)$$

We use the R programming language for all simulations and computations (R Core Team, 2016).

## 3. Results

### 3.1 Computation time

For all simulation studies for the three copula models above, we show average computation time for copula parameter estimation in Table 1, for both the subset analyses and the full data analyses. For the Gaussian copula with full data sample size 50,000, the average computing time for 10 subsets is 63 times faster than the full data analysis; this assumes that all jobs are run in parallel at the same time. The average computing time decreases as the number of subsets increases, with 100 subsets analyzed 1,171 times faster than the full data set. We also found that as the full data sample size increases, the average computing time is reduced further in parallel computing. For example, when sample size is quadrupled to 200,000, the average computing time for 10 subsets is 112 times faster than the full data set, and for 100 subsets, it is 3,607 times faster. We also found appreciable improvements in average computation time for the Frank and Gumbel copulas (Table 1), with average computing time 3,086 times faster for 100 subsets versus the full data analysis for the Frank copula, for sample size 200,000, and 1,682 times faster for the Gumbel copula.

## 3.2 Estimated bias, estimated MSE, and estimated relative $L_1$ and $L_2$ errors

For the unknown $\theta$ parameter of the Gaussian, Frank and Gumbel copulas, the results for the estimated bias, estimated MSE, and estimated relative $L_1$ and $L_2$ errors are shown in Figures 1, 2 and 3, respectively, for all simulation studies (see definitions of these values in Section 2). The estimated bias, estimated MSE, and estimated relative $L_1$ and $L_2$ errors increase as the number of subsets increases; and all of these values decrease as the sample size increases. For the Gaussian copula, the estimated bias is less than 0.002 for all numbers of subsets and all sample sizes. The estimated bias is smallest for sample size 200,000 and 10 subsets, where the estimated bias is $3.8 \times 10^{-5}$ (Figure 1). For the Frank copula, the estimated bias values are larger than for the Gaussian copula. Here, the estimated bias is less than 0.008 for all simulation studies; it is again smallest for sample size 200,000 and 10 subsets, where it is 0.0002. For the Gumbel copula, the estimated bias values are also larger than for the Gaussian copula. The estimated bias is less than 0.008 for all simulation studies; it is also smallest for sample size 200,000 and 10 subsets, where it is 0.0003. For estimated MSE, for all simulation studies, these values are less than $4\times10^{-6}$ for the Gaussian copula, less than $7\times10^{-5}$ for the Frank copula and less than $9\times10^{-5}$ for the Gumbel copula. For the estimated relative $L_1$ and $L_2$ errors, these values are lower than 0.0013 for all simulation studies and all copulas (see Figures 1, 2, and 3).

## 4 Discussion

We found in simulation studies that our new method using parallel computing greatly reduces computation time for estimating copula parameters, with three commonly-used one-parameter bivariate copula models. The computation time was between 45 and 3,607 times faster for parallel computing versus full data analysis, depending on the number of subsets, sample size, and copula model used. The parameter estimates from our new method showed small estimated bias, with values less than 0.008 for all simulation studies and copulas. We also found small values for estimated MSE, estimated relative $L_1$ errors, and estimated relative $L_2$ errors, across all simulation studies and copulas. Of the three copulas studied, the Gaussian copula had the smallest values for estimated bias and estimated MSE for all simulation studies.

The choice of the number of subsets is typically based on the number of machines or processors available to a user. We found in simulation studies that as the number of subsets increases, the estimated bias increases, but the average computing time decreases. Thus, the user can determine the tradeoff between computing time and estimated bias in the results; note that the estimated bias is still reasonable, even for 100 subsets. Once the number of subsets is chosen in our simulation studies, the full data set is partitioned evenly across subsets. In practice, the user can choose to partition the full data set with different subset sample sizes. Since our subset combining method uses a weighted average of subset estimates, with weights based on the inverses of the asymptotic variances of the parameter estimates, unequal subset sample sizes are taken into account in the weighting. We plan to examine in future research the effects of methods of data partitioning on parameter estimation.

Our subset combining method uses embarrassingly parallel subset analyses, with the goal of speeding copula parameter estimation for big data. Our method can also be used for data sets that are required to be analyzed in parts; this can occur when data sets are too large to read into computer memory in full, when data sets have multiple sources, or when data sets are stored across numerous machines. We plan to extend our research to copula models with more than one parameter, as well as to multivariate copulas with dimension of three and higher.

25. Kojadinovic, I & Yan, J (2010), 'Comparison of three semiparametric methods for estimating dependence parameters in copula models', *Insurance: Mathematics and Economics*, **47**, 52-63.

26. Malevergne, Y & Sornette, D (2003), 'Testing the Gaussian copula hypothesis for financial assets dependences', *Quantitative Finance*, **3**, 231-250.

27. Mighell, KJ (2010), 'CRBLASTER: A parallel-processing computational framework for embarrassingly parallel image-analysis algorithms', *Publications of the Astronomical Society of the Pacific*, **122**, 1236-1245.

28. Neiswanger, W, Wang, C & Xing, E (2014), 'Asymptotically exact, embarrassingly parallel MCMC', In *Proceedings of the Thirtieth Conference on Uncertainty in Artificial Intelligence*, Quebec City, Quebec, Canada, 623-632.

29. Nelsen, RB (2006), *An Introduction to Copulas*, *Springer*, New York.

30. Nikoloulopoulos, AK, Joe, H & Li, H (2012), 'Vine copulas with asymmetric tail dependence and applications to financial return data', *Computational Statistics and Data Analysis*, **56**, 3659-3673.

31. Oakes, D (1994), 'Multivariate survival distributions', *Journal of Nonparametric Statistics*, **3**, 343–354.

32. Oliva, J, Póczos, B & Schneider, J (2013), 'Distribution to distribution regression', In *Proceedings of the 30th International Conference on Machine Learning, Journal of Machine Learning Research Workshop and Conference Proceedings*, Brookline, MA, 1049-1057.

**Tables**

**Table 1.** Average computation time for copula parameter estimation

| Copula Model | Sample Size | Average Time Per Subset | | | |
|---|---|---|---|---|---|
| | | 10 Subsets | 20 Subsets | 100 Subsets | Full Data |
| Gaussian | 50,000 | 1.11 | 0.36 | 0.06 | 70.27 |
| | 100,000 | 4.08 | 1.35 | 0.17 | 362.16 |
| | 200,000 | 10.92 | 3.66 | 0.34 | 1226.25 |
| Frank | 50,000 | 1.52 | 0.57 | 0.10 | 101.21 |
| | 100,000 | 3.99 | 1.25 | 0.17 | 313.84 |
| | 200,000 | 10.98 | 3.38 | 0.35 | 1079.94 |
| Gumbel | 50,000 | 1.88 | 0.76 | 0.14 | 84.58 |
| | 100,000 | 4.63 | 1.65 | 0.24 | 251.96 |
| | 200,000 | 12.56 | 4.60 | 0.61 | 1026.08 |

Average computational time, in seconds, for estimating the copula parameters. The results are averaged over simulation studies for the full data values, and averaged over subsets and simulation studies for the subset values. A computer with operating system Linux 3.13.0 and an Intel Xeon Processor E3-1225 V2 (8M Cache, 3.2 GHz) was used for all analyses.

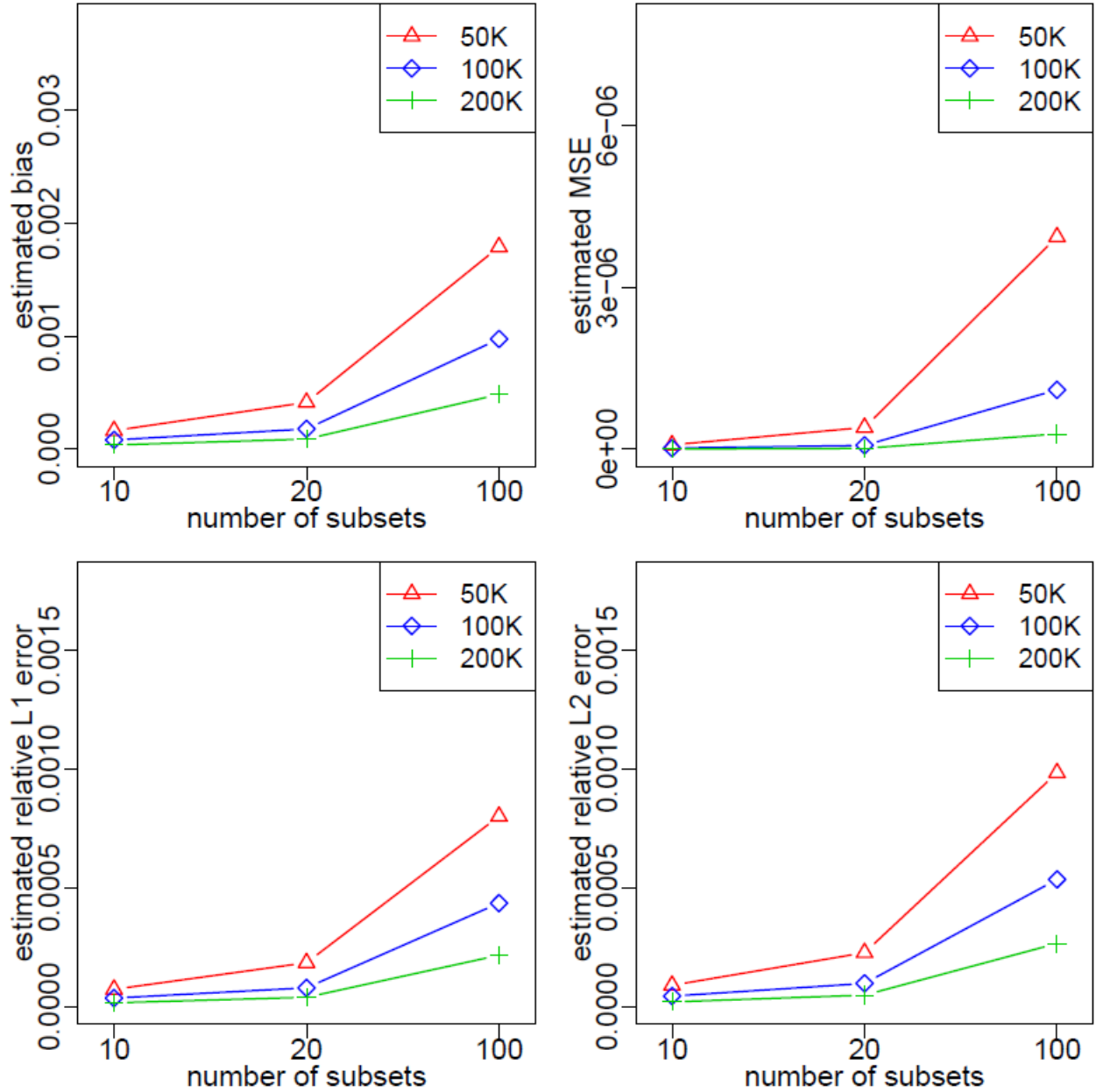

**Figure 1.** Estimated bias (upper left), estimated MSE (upper right), estimated relative $L_1$ error (lower left) and estimated relative $L_2$ error (lower right) for the $\theta$ parameter of the Gaussian copula. Results are shown for number of subsets = 10, 20, 100 and sample size = 50,000; 100,000; 200,000.

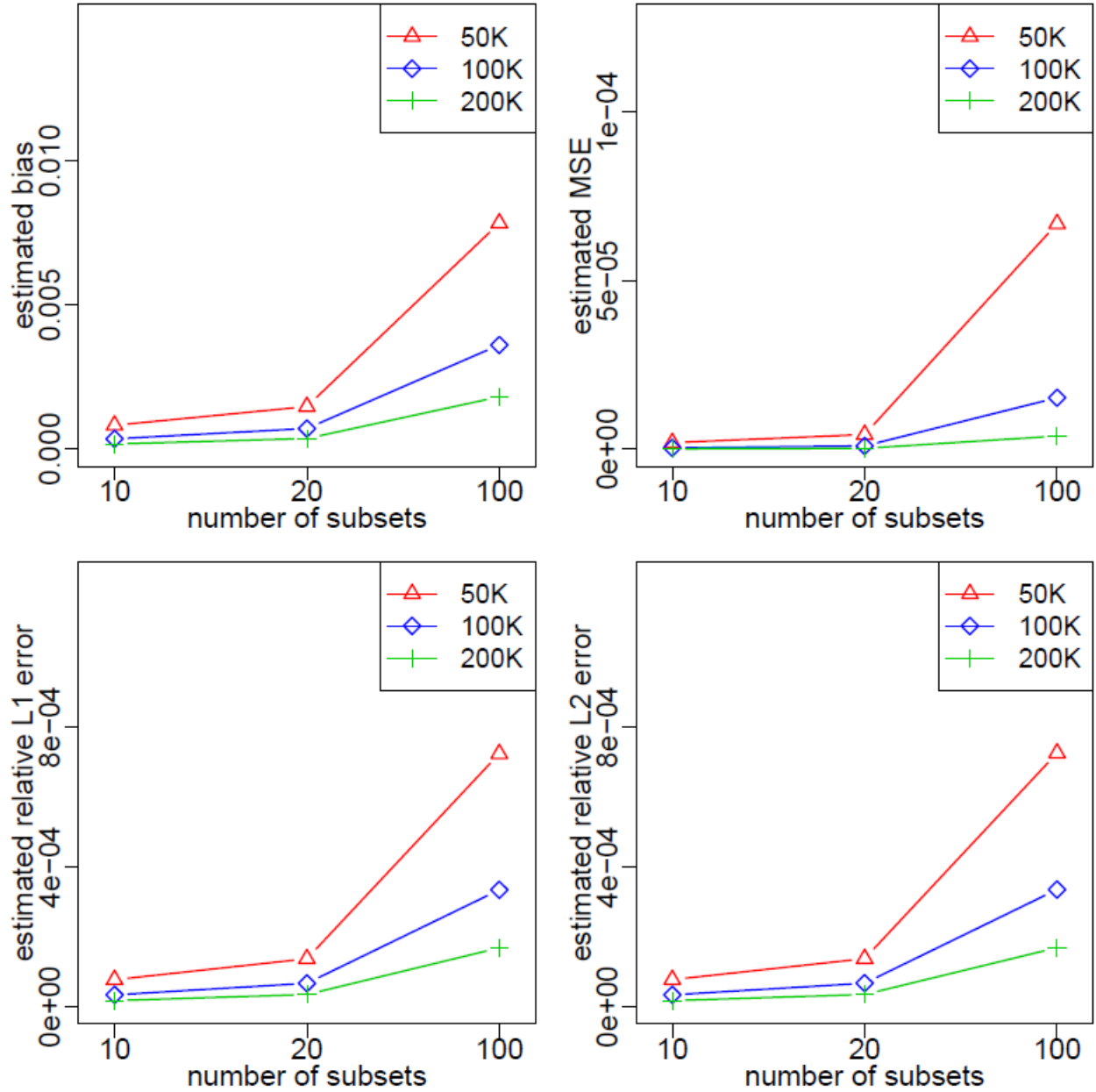

**Figure 2.** Estimated bias (upper left), estimated MSE (upper right), estimated relative $L_1$ error (lower left) and estimated relative $L_2$ error (lower right) for the $\theta$ parameter of the Frank copula. Results are shown for number of subsets = 10, 20, 100 and sample size = 50,000; 100,000; 200,000.

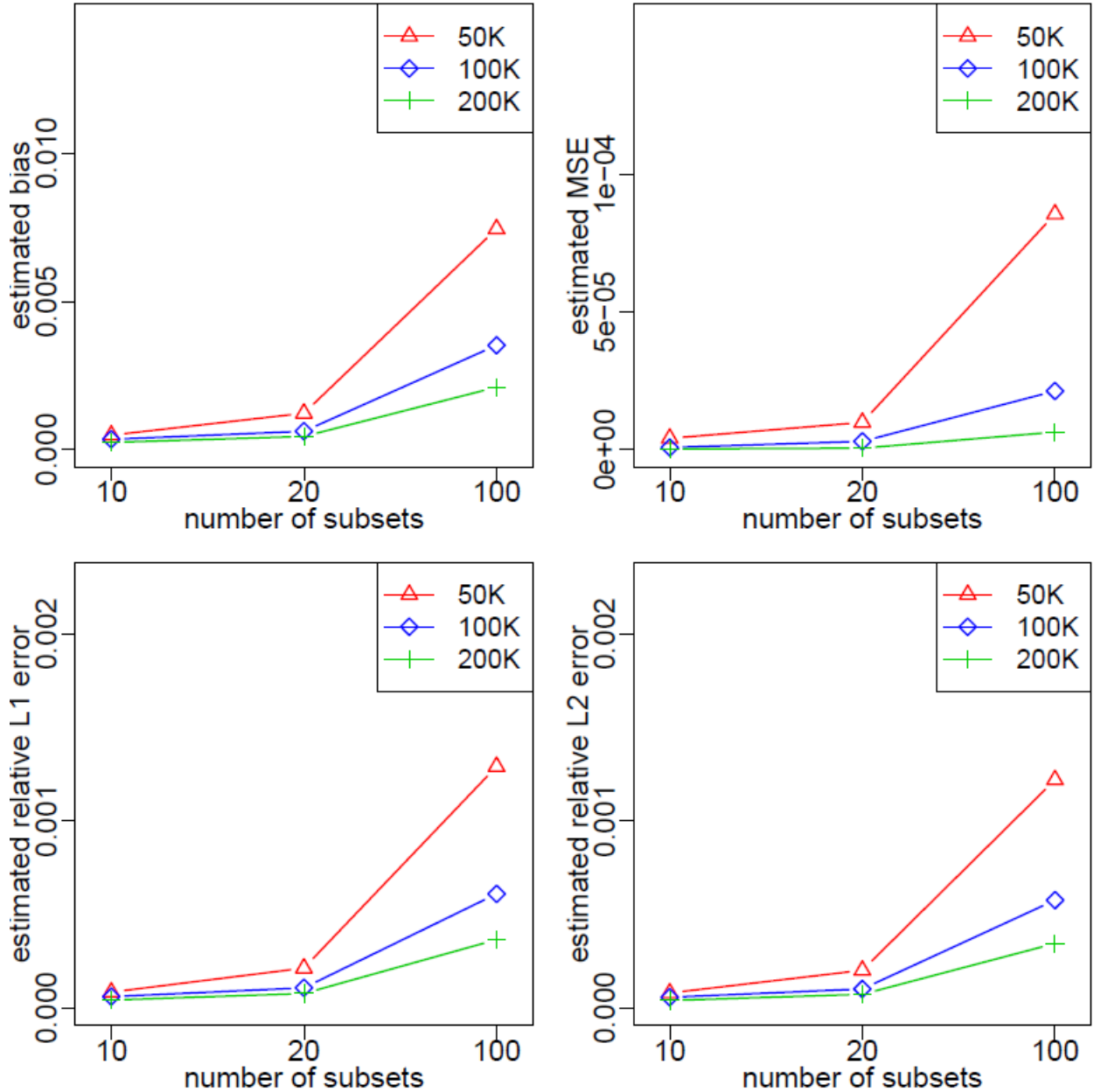

**Figure 3.** Estimated bias (upper left), estimated MSE (upper right), estimated relative $L_1$ error (lower left) and estimated relative $L_2$ error (lower right) for the $\theta$ parameter of the Gumbel copula. Results are shown for number of subsets = 10, 20, 100 and sample size = 50,000; 100,000; 200,000.